\documentclass[twocolumn,showpacs,prb]{revtex4}

\usepackage{amssymb}
\usepackage{amsmath}
\usepackage{epsf}
\usepackage{graphicx}
\usepackage[dvips]{epsfig}
\usepackage{bm}

\begin{document}
\title{Analysis of multiwalled carbon nanotubes as waveguides and antennas in the infrared and the visible regimes}

\author{M. V. Shuba, G. Ya. Slepyan, S. A. Maksimenko}
\affiliation{\\ Institute for Nuclear Problems, Belarus State
University, Bobruiskaya 11, 220050 Minsk, Belarus}
\author{C. Thomsen}
\affiliation{\\ Institut f\"{u}r Festk\"{o}rperphusik, Technische
Universit\"{a}t Berlin, Hardenbergstr. 36, D-10623 Berlin,
Germany}
\author{A. Lakhtakia}
\affiliation{\\
Nanoengineered Metamaterials Group, Department of Engineering Science and Mechanics,
Pennsylvania State University, University Park, PA 16802-6812,
USA}

\begin{abstract}

The propagation of azimuthally symmetric guided waves in
multiwalled carbon nanotubes (MW\-CNTs) was analyzed theoretically
in the mid-infrared and the visible regimes. The MWCNTs were
modeled as ensembles of concentric, cylindrical, conducting
shells. Slightly attenuated guided waves and antenna resonances
due to the edge effect exist for not-too-thick MWCNTs in the far-
and mid-infrared regimes.   Interband
transitions hinder the propagation of guided waves and have a
deleterious effect on the performance of a finite-length MWCNT as
an antenna. Propagation of surface-plasmon waves along an MWCNT
with a gold core was also analyzed. In the near-infrared and the
visible regimes, the shells behave effectively as lossy
dielectrics suppressing surface-plasmon-wave propagation along the
gold core.

\end{abstract}
 \maketitle

\section{Introduction}

Their unusual physical and chemical properties, and their
potential applications in a variety of nanotechnologies make
carbon nanotubes (CNTs) very interesting objects to
technoscientists, \cite{Dressel_B01,Reich_b04} despite possible
health hazards. \cite{Health} In particular,  CNTs have been
proposed to fabricate several different integrated-circuits
elements and electromagnetic devices, such as transmission lines,
\cite{Slepyan99,Maksim00,Maksim1,Hagmann_05,Rybczynski_07_APL}
interconnects,
\cite{Raychowdhury_06,Chiariello_07,Maffucci_08,Li_08} and
nanoantennas.
\cite{Wang_04_APL,Hanson05,Slepyan06,Burke06,Kempa_07,Wang_08_IJIRMW,Maksimenko07,Hanson08}
The fabrications of a  CNT-based amplitude modulator/demodulator
\cite{Rutherglen} and a fully integrated radio receiver
\cite{Iensen} have been reported. References
\onlinecite{Misewich_03_Sci,Chen_05_Sci,Kibis3,Kibis_07_NL,Batrakov_06_SPIE,Kuzhir_07_SRIMN,Batrakov_08_PhE}
demonstrate the potential of CNTs as emitters of terahertz and
infrared radiation.

Not surprisingly, the electromagnetic characteristics of CNT-based antennas
have been examined in different
frequency regimes ranging from the microwave to the visible.   CNT morphology
has been demonstrated to play a crucial role, as evinced by reported research
on  single-wall CNTs (SWCNTs),
\cite{Hanson05,Slepyan06,Burke06,Miano_06} $\sim 1$-cm-long
one-dimensional chains of electrically connected SWCNTs, \cite{Burke06}  planar
periodic structures of SWCNTs,\cite{Hanson06,Hanson07}  CNT
bundles \cite{Shuba07,Huang_08}, and CNT arrays.\cite{Lan_06} Continuing
in that vein, here we report our work on the performance of
 multiwalled CNTs (MWCNTs) as antennas.

A multiwalled CNT comprises $N$
concentric shells (or tubes), each obtained by rolling a
graphene sheet into a cylinder. The number of shells in a MWCNT can
range from 2 to 200, and the distance between consecutive shells
from $3.4$ to $3.6$ {\AA},\cite{Bandow}  which is close to the interlayer
distance in graphite ($3.35$ {\AA}). The lattice structures of
consecutive shells are generally uncorrelated with each other, and
can even have different chiralities. In fact, several experiments on
MWCNTs have indicated that often the different shells have different
periodicities.\cite{Iijima,Ge}  Two consecutive shells of an MWCNT are
called commensurate (incommensurate) if the ratio of their
unit-cell lengths along the CNT axis is rational
(irrational), indicating the presence (absence) of translational symmetry.
Incommensurability affects the transport and optical properties of
MWCNTs.
\cite{wang05,Saito93,Lambin00,Ahn03,Yoon02,Lunde05,Ho,Kwon98}

The critical issue when modeling the electromagnetic
properties of an MWCNT  is the
intershell interaction leading to intershell electron tunneling or
hopping.  Published data, although very variable, shows a strong
dependence on the  intrinsic symmetries of the shells, which dictates
selection rules for the elements of the tunneling matrix, as determined by the conservation
laws for
energy and momentum. As may be expected, two incommensurate shells
interact differently than two commensurate cells.
\cite{wang05,Saito93,Lambin00,Ahn03,Yoon02,Lunde05,Ho,Kwon98} For
example,  the Fermi momentums of two incommensurate shells do not
coincide within the first Brillouin zone and therefore the
intershell tunneling vanishes. \cite{Yoon02}

An isolated MWCNT can be modeled in several
different ways. Abrikosov
\emph{et al.} \cite{Abrikosov} considered an MWCNT as a set of coaxial,
continuous, conducting cylinders accompanied by an appropriate Kronig-Penney-type
potential in the radial direction. Dyachkov and Makaev \cite{Dyachkov}
as well as Tunney and Cooper \cite{Tunney} assumed
the intershell interaction   to be so small that each shell could be considered to
be in a perturbed eigenstate of an SWCNT. A computer simulation with
some input parameters extracted from
experiment has also been reported.\cite{Bourlon}

Following
Refs. \onlinecite{Dyachkov} and  \onlinecite{Tunney}, we decided to neglect the contribution of the
intershell tunneling to the radiation characteristics of an MWCNT. Indeed,   though intershell interaction in
defectless finite-length double-wall CNTs shifts optical band gaps and
distorts the density of electronic states, it does not subvert the
intrinsic type of conductivity (either metallic or semiconductor);\cite{Abrikosov,Dyachkov,Tunney, Bourlon}
therefore, it would influence the radiation characteristics only slightly.
As the presence  of localized defects can dramatically increase
intershell conduction, \cite{Tunney}  we restrict ourselves here to MWCNTs with
defect-free shells by neglecting intershell
conduction.

Analysis of the physical characteristics of CNT-based integrated
circuits elements, such as antennas,   must follow the general
principles of electrodynamics and must account for the peculiar
dispersion properties of electrons in CNTs. A key element of the
analysis is the  formulation of the effective boundary conditions for the
electromagnetic field on the CNT surface.
\cite{Slepyan99,Maksim00,Maksim1} These boundary conditions are
dictated by the microscopic model of CNT used. Hence, any attempt
 to describe an MWCNT with cross-sectional radius between $25$ and $150$~nm as a
perfectly conducting rod \cite{Wang_04_APL,Lan_06}
can provide only
 rough estimates, and is inapplicable to
MWCNTs of small radius.

We adopt a proper microscopic model for   MWCNTs here. A relevant boundary-value problem for an
MWCNT of finite length is formulated in Sec.~\ref{model}.
Since the radiation characteristics of an MWCNT are
determined by its waveguiding properties, the dispersion equation
for guided-wave propagation on an infinitely long MWCNT is derived in
Sec.~\ref{guided}.
The boundary-value problem of Sec.~\ref{model}
is solved
numerically by an integral-equation technique in Sec.~\ref{finite-length}. This approach is well
established in
antenna theory \cite{balanis} and has been successfully applied to
SWCNT antennas \cite{Hanson05} and almost circular bundles of closely packed SWCNTs.
\cite{Shuba07} Section~\ref{Num} contains numerical results for
guided-wave parameters (slow-wave and attenuation) and the scattering
properties of MWCNT in a wide frequency range from the terahertz to the
visible regimes. Sections~\ref{Discus}  presents an assessment of the potential of an isolated MWCNT
 function as an
optical nanoantenna. Concluding remarks are provided
in Sec.~\ref{concl}.  An $\exp(-i\omega t)$ time-dependence
is implicit, $t$ denotes the time, and $\omega$ is the angular
frequency.

\section{Model} \label{model}

Let us model an MWCNT as a  multishell structure
comprising $N$ coaxial, infinitesimally thin cylinders in free space. Each shell is
 achiral, i.e., it has either a zigzag ($m,0$) or an
armchair ($m,m$) configuration, with $m$ as an integer. Therefore,
our model is applicable to both semiconducting and metallic shells.
The sequence of armchair and zigzag shells in the MWCNT is
arbitrary.

Let the cylindrical axis of the chosen MWCNT be parallel to the $z$ axis
of the
cylindrical coordinate system $\left(\rho, \phi, z\right)$. The centroid
of the MWCNT is assumed to coincide with the origin of the coordinate system.

Let us enumerate  shells in the MWCNT consecutively from $1$ to
$N$ beginning from the innermost shell, so that their cross-sectional
radii comply with the condition $R_N > R_{N - 1}
> ... > R_1$.     The cross-sectional radius $R_N$ of  the outermost shell
is assumed to be much
smaller than the free-space wavelength.
 The $p$-{th}
shell, $p\in\left[1,N\right]$, possesses an axial conductivity per unit area denoted by
$\sigma_p$.  The transverse current density
in every shell is neglected.
Intershell tunneling is also neglected, as mentioned in the previous section.

As electromagnetic fields with azimuthal symmetry are easily
excited in an MWCNT by a uniform external field, we decided to
restrict ourselves to azimuthally symmetric fields. This
restriction also holds for finite-length MWCNTs in the
long-wavelength regime (Sec. \ref{num1}). The electric Hertz
vector ${\rm {\bf \Pi }}\equiv \Pi(\rho,z) {\rm {\bf e}}_z $ is
then governed by the Helmholtz equation
\begin{equation}
\label{eq1} \frac{1}{\rho}\frac{\partial}{\partial \rho}
\Bigl(\rho{\frac{\partial \Pi}{\partial \rho}} \Bigr)+
\frac{\partial ^2 \Pi}{\partial z^2} + k^2\Pi = 0 \,,
\end{equation}
where ${\rm{\bf e}}_z $ is the unit vector along the $z$ axis, $k
= \omega / c$ the free-space wavenumber,  and $c$ is speed of light in free
space (i.e., vacuum). Since $ \Pi$
depends only on $\rho$ and $z$,  the components of the
electric and magnetic fields  are as
follows:
\begin{equation}
\label{elfield1}
\begin{array}{lll} \displaystyle
E_{\rho} = \frac{\partial ^2\Pi}{\partial \rho\partial z}\,, &
E_\phi=0\,,& \displaystyle E_z =\Bigl( {\frac{\partial
^2}{\partial z^2} + k^2} \Bigr)\Pi\,,
\end{array}
\end{equation}
\begin{equation}
\label{elfield2}
\begin{array}{lll}
H_{\rho} = 0\,, &\displaystyle H_\phi=ik\frac{\partial \Pi
}{\partial \rho }\,,& H_z =0\,.
\end{array}
\end{equation}

On neglecting intershell
tunneling, the two boundary conditions across the $p$-{th} shell are given by \cite{Slepyan99}
\begin{eqnarray}
&\label{boundary3} \displaystyle \left.{\frac{\partial \Pi
}{\partial \rho}}
    \right|_{\rho= R_p +0} -\left. {\frac{\partial \Pi }{\partial \rho}}
    \right|_{\rho = R_p - 0} = \frac{4\pi}{ikc} J_p\,,
    \\ \rule{0in}{4ex}
    & \Bigl.\Pi \Bigr|_{\rho = R_p + 0} =  \Pi\Bigr| _{\rho = R_p -
    0}\,. \label{boundary4}
\end{eqnarray}
Here ${\bf J}_p(z)=J_p(z) {\bf e}_z$ is the axial current density on
the surface $\rho = R_p$, with
\begin{equation} \label{current}
J_p(z) = \sigma_p
     \left( \frac{d^2}{d z^2} +
    k^2\right) \Pi (R_p, z)  \,.
\end{equation}

The surface conductivity $\sigma _p $ of the $p$-{th} isolated shell  is available via quantum-transport calculations carried out in the tight-binding approximation as
\cite{Slepyan99,MKSK}
\begin{widetext}
\begin{equation}
\label{eq4} \sigma _p (\omega )=  - \frac{ie^2\omega }{\pi
^2\hbar R_p}\left\{
    \frac{1}{\omega(\omega + i/\tau)}\sum\limits_{s = 1~}^{m}{\int\limits_{1stBZ}\!\!
    { \frac{\partial F_c}{\partial p_z}
    \frac{\partial {\cal E}_c}{\partial p_z }} }dp_z
    -2\sum\limits_{s = 1~}^{m} {\int\limits_{1stBZ}\!\! {\cal E}_c \left|
    R_{vc} \right|^2
    \frac{F_c - F_v}{\hbar^2\omega(\omega + i/\tau) - 4{\cal E}_c ^2}}dp_z \right\}
\,,
\end{equation}
\end{widetext}
where $e$ is the electron charge, $\hbar $ is the normalized
Planck constant, and $p_z $ is the axial projection of
quasi-momentum of an electron. The integer $s \in\left[1,m\right]$
labels the $\pi$-electron energy bands and the abbreviation
$1stBZ$ restricts the variable $p_z $ to the first Brillouin zone.
The time constant $\tau$ (being the relaxation time) is assumed to
be equal to the inverse of the relaxation frequency $\nu$. The
normalized matrix elements of the dipole transition between the
conduction and valence bands, denoted by $R_{vc}$, are evaluated
in the tight-binding approximation, after taking into account the
transverse quantization of the charge carriers' motion and the
hexagonal structure of the graphene lattice. \cite{Maksim1}

The first term on the right side of (\ref{eq4}) is the Drude term
corresponding to the intraband conductivity, while the second term
describes the contribution of interband transitions between the
valence and the conduction bands.
 The indexes $c$ and $v$  refer to the
conduction  and valence bands, respectively. The equilibrium Fermi
distribution functions
\begin{equation}
F_c(p_z ,s)=
    \left\{ 1+\exp\left[\displaystyle{
 \frac{{\cal E}_{c}(p_z,s)} {k_B T}
     } \right] \right\} ^{-1}
    \end{equation}
    and
    \begin{equation}
F_{\upsilon}(p_z ,s)=
    \left\{ 1+\exp\left[\displaystyle{
 \frac{{\cal E}_{\upsilon}(p_z,s)} {k_B T}
     } \right] \right\} ^{-1}
    \end{equation}
involve the temperature  $T$ and the Boltzmann constant $k_B $. The
electron energies ${\cal E}_{c,\upsilon} (p_z ,s)$ and normalized matrix elements
of the dipole transition  $R_{vc}$ for zigzag ($m,0$) CNT are given by
\cite{Dressel_B01}
\begin{eqnarray}
\nonumber
 {\cal E}_{c}(p_z,s) &=&-  {\cal E}_{\upsilon}(p_z,s)
 \\
 \nonumber
&=&\gamma _0 \,\sqrt {1 + 4\cos
\left( {ap_z } \right)\cos \left( {\frac{\pi
s}{m}} \right) + 4\cos ^2 \left( {\frac{\pi s}{m}} \right)}\,
\\
&&
\label{dispers1}
\end{eqnarray}
and
\begin{eqnarray}
\nonumber
 &&R_{vc} (p_z,s)= -\frac{b \gamma_0^2}{2{\cal E}_{c}^2} \\
 &&\quad\times\left[ 1+\cos(ap_z)
\cos \left( \frac{\pi s}{m} \right) - 2\cos^2 \left( \frac{\pi
s}{m} \right)\right]\,,
\end{eqnarray}
respectively, where $\gamma _0 \approx 2.7$ eV is the overlap
integral, \cite{Dressel_B01} $a = 3b / 2\hbar$, and $b=0.142$~nm is
the interatomic distance in graphene. Expressions for ${\cal
E}_c(p_z ,s)$ and $R_{vc}(p_z ,s)$ for armchair ($m,m$) CNTs are
available elsewhere. \cite{Dressel_B01}

\section{Guided-wave propagation in an infinitely long MWCNT} \label{guided}

Let us recall that intershell tunneling is negligibly small so that
$\Pi$ can be represented as a superposition of $N$ independent
functions; furthermore, we seek a solution of (\ref{eq1}) in the
form of a guided wave. Accordingly,
\begin{equation} \label{eq5-5}
 \Pi (\rho ,z ) = e^{ihz}{\sum\limits_{p = 1}^{N} }
A_p\,\Phi_p(\rho) \,,
 \end{equation}
where $\left\{A_p\right\}$ is the set of unknown coefficients and
$h$ is the unknown guide wavenumber. With $\kappa = \sqrt {h^2 -
k^2} $, the basis function $\Phi_p(\rho)$ is taken to
satisfy the differential equation
\begin{equation}
\label{eq1-2}
    \frac{1}{\rho}\,\frac{d}{d\rho}\left[\rho\frac{d}{d\rho}\,\Phi_p(\rho)\right]
     + \kappa^2 \Phi_p(\rho) =
    0\,, \quad p\in[1,N]\,,
\end{equation}
subject to the following boundary conditions at the surface $\rho = R_p$:
\begin{eqnarray}
&\label{boundary3-1} \displaystyle \left.{\frac{\partial
}{\partial \rho}}\Phi_p(\rho)
    \right|_{\rho= R_p +0} -\left. {\frac{\partial }{\partial \rho}} \Phi_p(\rho)
    \right|_{\rho = R_p - 0} = \frac{4\pi}{ikc} \,,
    \\ \rule{0in}{4ex}
    & \Bigl.\Phi_p(\rho) \Bigr|_{\rho = R_p + 0} =  \Phi_p(\rho)\Bigr| _{\rho = R_p -
    0}\,. \label{boundary4-1}
\end{eqnarray}
The appropriate solution of \eqref{eq1-2}  is
\begin{equation} \label{eq5-6}
\Phi_p (\rho)= \frac{4i\pi R_p}{kc} \left\{ {\begin{array}{ll}
 K_0 (\kappa R_p )I_0 (\kappa \rho )\,,\quad\rho < R_p \,, \\ \rule{0in}{3ex}
 I_0 (\kappa R_p )K_0 (\kappa \rho )\,,\quad\rho > R_p \,, \\
 \end{array}} \right.
\end{equation}
where $I_0 ( \cdot )$ and $K_0 ( \cdot )$
are modified Bessel functions of the zeroth order.

Whereas the boundary condition (\ref{boundary4}) is automatically satisfied by (\ref{eq5-5}) and (\ref{eq5-6}),
(\ref{boundary3})  remains to be satisfied. Substitution of (\ref{eq5-5}) into
(\ref{boundary3}) leads to a system of $N$ linear homogeneous equations
with respect to the coefficients $A_p$. A
nontrivial solution of the system
is provided by the dispersion equation
\begin{equation}
\label{eq12} \det \textsf{M} = 0\,.
\end{equation}
The elements $M_{qp}$ of the $N\times N$ matrix $\textsf{M}$ are given by
\begin{eqnarray}
\begin{array}{ll}
  {\displaystyle  M_{qp} =  \left\{
  {\begin{array}{l}
 K_0 (\kappa R_q )I_0 (\kappa R_p ), \quad q > p \,, \\ \rule{0in}{4ex}
  K_0 (\kappa R_p )I_0
 (\kappa R_q ) -
{\displaystyle \frac{i\omega \delta_{qp} }{4\pi R_q \sigma_q \kappa ^2}} \,, \quad q \leq p \,, \\
 \end{array}} \right.}
\end{array}
 \end{eqnarray}
where $\delta_{qp}$ is the Kronecker delta.  The dispersion equation
(\ref{eq12}) has $N$ $\kappa$-roots, each corresponding to a guided wave in the
MWCNT. From each $\kappa$-root, we can determine $h$ and
 the
slow-wave coefficient $\beta = k/h$.

\section{Light scattering by a finite-length MWCNT}
 \label{finite-length}

 \subsection{Integral-equation
technique} \label{inteq}
Let us now consider the scattering of an
electromagnetic wave incident on a   MWCNT of finite length $L$. The scattered
field can be described by (\ref{elfield1}) and (\ref{elfield2})
with the electric Hertz vector satisfying the Helmholz equation (\ref{eq1}), the usual
radiation condition, \cite{balanis} and the boundary conditions
(\ref{boundary3}) and (\ref{boundary4}). The surface current
density $J_p(z)$ at $\rho =R_p$ is given by
\begin{equation}
\label{eqJ} J_p(z)= \sigma_p
     \left( \frac{d^2}{d z^2} +
    k^2\right) \Pi (R_p, z)+\sigma_p E^{(0)}_z(R_p,z)  \,,
\end{equation}
where $E^{(0)}_z(\rho,z)$ is the $z$-component of the incident electric field.
As the intershell tunneling through the two ends of the MWCNT is negligible, the current density $J_p(z)$
satisfies the edge conditions
\begin{equation}
\label{eq14a} J_p (\pm L/2) = 0\,,
\end{equation}
thereby expressing the absence of concentrated charges on the  ends.
The boundary-value problem can effectively be solved by the integral-equation
technique for the surface current density as follows. \cite{Ilyinsky,Novotny}

The potential $\Pi (\rho,z )$ must be linearly related to $J_p(z)$, $p\in\left[1,N\right]$, as
\begin{equation}
\label{eq15a} \Pi (\rho,z ) = \frac{i}{\omega }\sum\limits_{p =
1}^{N} {R_p \int\limits_{ - L / 2}^{L / 2} {J_p ({z}')\,G (z -
{z}',\rho ,R_p )d{z}'} } \,,
\end{equation}
where
\begin{equation}
\label{eq16a} G(z,\rho ,R) = \int\limits_{0}^{2\pi} {\frac{\exp
\left( {ik\sqrt {\rho ^2 + R^2 - 2R\rho \cos \varphi + z^2} }
\right)}{\sqrt {\rho ^2 + R^2 - 2R\rho \cos \varphi + z^2}
}d\varphi }\,
\end{equation}
is the free-space Green function for (\ref{eq1}).
Setting $\rho = R_q $ in   (\ref{eq15a}) and making use of
(\ref{eqJ}), we arrive at the following system of $N$ integral equations with
respect to unknown current densities:
\begin{eqnarray}
 && \nonumber
 {\displaystyle    - \frac{1}{2ik}\int\limits_{
- L / 2}^{L / 2} {E^{(0)}_{z} ({R_q,z}')\exp (ik\vert z -
{z}'\vert
)d{z}'} }\\[5pt]
 && \label{eq18a}
 {\displaystyle\quad + \,C_q \exp (ikz) + D_q \exp ( - ikz)}
 \\
 &&\nonumber =
\sum\limits_{p = 1}^{N} {\int\limits_{ - L / 2}^{L / 2} { \left[
{\frac{2\pi i R_p }{\omega }G(z - {z}',R_p
,R_q )} \right.} }\\[5pt]
 && \qquad +
 \left. {\frac{i\delta _{qp} }{2k \sigma_p }\exp (ik\vert
z - {z}'\vert )} \right]
 J_p ({z}')d{z}'\,,\quad
q\in[1,N]\,. \nonumber
\end{eqnarray}
Here,
 $C_q $ and $D_q $ are unknown constants to be
determined from the edge conditions (\ref{eq14a}).

In the long-wavelength regime ($\lambda \gg L$), the
electromagnetic scattering properties of the MWCNT can be encapsulated in a
polarizability tensor with only one nonzero component
\begin{equation}
\label{eq20} \alpha_{zz} = \frac{2\pi i}{\omega E^{(0)}_{z}(0,0)
}\sum\limits_{p = 1}^N {R_p \int_{ - L / 2}^{L / 2} {J_p (z)dz} }\,.
\end{equation}
The integral on the right side of  (\ref{eq18a}) can be
numerically handled by a quadrature formula. \cite{Colton}
Parenthetically, the foregoing formalism was recently applied to almost
circular, closely packed bundles of finite-length parallel,
identical, metallic SWCNTs. \cite{Shuba07}

A finite-length MWCNT functioning as a receiving antenna
 can be characterized by the antenna efficiency \cite{Shuba07}
\begin{equation}
\label{eq21} \eta = \frac{P_r }{P_t + P_r }\,,
\end{equation}
\noindent where
\begin{equation}
\label{eq21a} P_r = \frac{\pi^2 k^2 }{c}{\int\limits_0^\pi
\sin^3\theta
 \left|
  { \int_{ -
L / 2}^{L / 2}  { e^{ikz\cos \theta }} {\sum\limits_{p = 1}^{N}
R_p J_p (z)\, dz } } \right|^2 d\theta}
\end{equation}
 is the scattered power and
\begin{equation}
\label{eq21b} P_t = \pi  {\sum\limits_{p = 1}^{N } {R_p\, {\rm Re}
\left( \frac{1}{\sigma _p } \right)\int_{ - L / 2}^{L / 2} {\vert
J_p (z)\vert ^2\,dz} } }
\end{equation}
is the power lost to ohmic dissipation.

 \subsection{Interband-transitions regime: first Born approximation} \label{low-coupling}

According to Refs.~\onlinecite{Slepyan06} and
\onlinecite{Hanson07}, surface waves in a CNT are strongly
attenuated  in the frequency regime of interband transitions.
Therefore, the surface current density $J(z)$ in an SWCNT of
radius $R$ exposed to an incident electric field
$\mathbf{E}^{(0)}$ obeys Ohm's law
\begin{equation}
\label{Ohm0}
J(z) = \sigma E^{(0)}_z(R,z)
\end{equation}
very well,
$\sigma$
being the surface conductivity.
Equation~(\ref{Ohm0}) can be considered as
 the first Born
approximation. \cite{Wolf}

Application of the first Born approximation (\ref{Ohm0}) to an MWCNT allows us
to express the axial surface current density $J_p(z)$ on  the surface
of the $p$-{th} shell as
\begin{equation}
\label{Ohm} J_p(z)\approx\sigma_p E^{(0)}_z (R_p,z)\,,
\quad p\in\left[1,N\right]\,.
\end{equation}
This expression can be justified as follows. The electric field exciting the $p$-th shell
is made of two components: (i) the electric field $\mathbf{E}^{(0)}$ incident on the
entire MWCNT and (ii) the sum of the electric fields radiated to the surface current densities
$J_q$, $q\in\left[1,N\right]$ but $q\ne p$. Equation~(\ref{Ohm}) is justified if
the first component is much larger than the second component
 at $\rho=R_p$, i.e., the condition
\begin{eqnarray}
\nonumber
&&\left| \Bigl(\frac{d^2}{d z^2} +
    k^2\Bigr) \Pi (R_p, z) \right| \ll |E^{(0)}_z(R_p,z)|\,,\\[5pt]
    &&\qquad\quad z\in\left[-L/2,L/2\right]\,,\quad
    p\in\left[1,N\right]\,,
    \label{Condition}
\end{eqnarray}
holds true.
Substituting (\ref{eq15a}) and (\ref{Ohm}) into
(\ref{Condition}), we arrive at the condition
\begin{eqnarray}
 \nonumber &&
 {\displaystyle  \left| \sum\limits_{q = 1}^{N} {R_q \sigma_q
\int_{ - L / 2}^{L / 2} {E^{(0)}_z(R_q,{z}')G (z - {z}',R_p
,R_q )d{z}'} }\right| }\\[5pt]
 && \nonumber
 {\displaystyle\quad \ll \left| \frac{c}{2}\int_{
- L / 2}^{L / 2} {E^{(0)}_{z} ({R_p,z}')\exp (ik\vert z - {z}'\vert
)d{z}'}\right|}\,,
\\[5pt]
    &&\qquad\quad z\in\left[-L/2,L/2\right]\,,\quad
    p\in\left[1,N\right]\,,
\label{condition1}
 \end{eqnarray}
for the applicability of the first Born approximation.  We found
that condition (\ref{condition1}) holds true for  MWCNTs that are
not too thick (i.e., $R_N$ is sufficiently small) in the frequency
regime wherein the surface conductivities $\sigma_p$,
$p\in\left[1,N\right]$, are mostly determined by interband
transitions --- the second term in (\ref{eq4}). We estimate that
MWCNTs with outermost radius $R_N < 20$~nm satisfy   (\ref{Ohm})
at frequencies in the mid-infrared and the visible regimes.

Equation~(\ref{Ohm}) contradicts
the edge conditions (\ref{eq14a}). An analogous situation appears,
for example, in the theory of diffraction by an aperture in an infinitesimally
thin, perfectly conducting screen, wherein the exact solution must satisfy the Meixner
condition \cite{Ilyinsky} on the aperture edge but an approximate
solution based on the Huygens principle does not satisfy that condition.
However, the error is strongly localized in the vicinity of the
edge of the aperture and does not influence the scattered field in the far zone.
 \cite{Born} For our problem, (\ref{Ohm}) may be
interpreted as a version of the Huygens principle for an
MWCNT: the scattered field is generated by secondary current densities
  induced by the  incident electric field on the surfaces of all shells.
Therefore, the applicability of   (\ref{Ohm}) is unphysical only
in the vicinity of the edges $z=\pm L/2$, but that should not affect
the performance of the MWCNT as an antenna in the interband-transitions
regime.

When the MWCNT is electrically thin in cross-section ($k R_N\ll 1$),
the further approximation
\begin{equation}
E^{(0)}_z(\rho,z)\simeq E^{(0)}_z(0,z)\,,\quad \rho \leq R_N
\end{equation}
is permissible. Then the substitution of (\ref{Ohm})
in  (\ref{eq21a}) and  (\ref{eq21b}) yields
\begin{equation}
\label{eq21c} P_r = \frac{\omega^2 \left|\sigma _{T}\right|^2}{4c^3}{\int_0^\pi
\sin^3\theta
 \left|
  { \int_{ -
L / 2}^{L / 2}  { e^{ikz\cos \theta }} E^{(0)}_z(0,z) dz} \right|^2
d\theta}\,,
\end{equation}
\begin{equation}
\label{eq21d} P_t = \frac{1}{2} {\rm Re}\left(\sigma _{T}\right) {\int_{ - L /
2}^{L / 2} {\vert E^{(0)}_z(0,z)\vert ^2dz} }\,,
\end{equation}
where
 \begin{equation} \label{sigm}
\sigma  _{T} = \sum\limits_{p = 1}^N \,\left({2\pi R_p \sigma _p }\right)
\end{equation}
is the {\it effective} conductivity per unit length   of an electrically thin MWCNT.
The condition (\ref{condition1}) leads to the
inequality $P_t\gg P_r$; consequently,
\begin{equation} \label{KPD}
\eta \approx P_r/P_t \sim \left|\sigma _{T}\right|^2/{\rm Re}\left(\sigma  _{T}\right) \,.
\end{equation}

As determined by (\ref{eq21c}) and (\ref{eq21d}), $P_t$ and $P_r$
are the same as for a thin-wire resistive antenna whose
conductivity per unit length is equal to $\sigma_{T}$.\cite{King}
 Thus, an electrically thin MWCNT which
satisfies the condition (\ref{condition1})  may be modeled as a
thin homogenous cylinder with conductivity per unit length
determined from (\ref{sigm}).

\section{Numerical results}
 \label{Num}

In order to present illustrative results, let us consider an MWCNT
consisting of only zigzag shells. A large number of such MWCNTs are
possible. \cite{Dressel_B01,Reich_b04} Moreover, after a suitable
modification of the effective boundary conditions
(\ref{boundary3}) and (\ref{boundary4}),\cite{MKSK} the
approach developed can be extended to MWCNTs comprising shells with arbitrary chirality
vectors.

\begin{figure}[!htb]
\begin{center}
\includegraphics[width=3.4in]{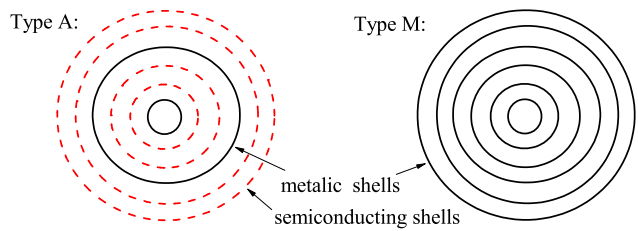}
\end{center}
\caption{Schematics of MWCNTs of types A and M. All shells in an
MWCNT of type M are metallic. In contrast, an MWCNT of type A contains a metallic shell
alternating with two semiconducting shells.
} \label{fig1}
\end{figure}

For definiteness, calculations were performed for two types of
MWCNTs, hereafter referred to as type A and type M, shown in Fig.
\ref{fig1}. The $p$-th shell in an MWCNT of type A is in the
$(8p+1, 0)$ configuration; hence, two consecutive semiconducting
shells are followed by a metallic shell. The $p$-th shell in an
MWCNT of type M is in the $(9p, 0)$ confuguration; hence, all $N$
shells are metallic. The radius of the $p$-th shell is given by
\begin{equation}
\label{Rp-def}
R_p =
\left\{
\begin{array}{ll}
\sqrt{3}(8p+1)b/(2\pi)\,, &\quad{\rm type\,\, A}\\[5pt]
9\sqrt{3}pb/(2\pi)\,, &\quad{\rm type\,\, M}
\end{array}\right.\,.
\end{equation}

 The frequency $f^{(1)}_p$ of the first interband transition
 for the $p$-{th} shell of the metallic type is determined by the energy difference
 between the lowest van Hove singularity of the conducting band
and the highest  van Hove singularity of the valence band. Thus,
an analysis of the dispersion equation (\ref{dispers1}) for zigzag
shells yields
\begin{eqnarray}
\nonumber
f^{(1)}_p &=& 2{\cal E}_{c}[2\pi/(3a),3p-1]/(2\pi\hbar)
\\
&\simeq& \upsilon _F / (\pi R_p
 )\,,
 \end{eqnarray}
 where $\upsilon _F$ is the $\pi$-electron velocity
 at the Fermi level. The frequency
of the first interband transition for the $p$-{th} shell of the
semiconducting type is determined by the minimal band-gap energy as
\begin{eqnarray}
\nonumber
f^{(1)}_p &=& 2{\cal E}_{c}[2\pi/(3a),3+[8(p-1)\pm1]/3]/(2\pi\hbar)\\
&\simeq& \upsilon _F / (3\pi R_p )\,,
\end{eqnarray}
wherein the sign have to be chosen so that $[8(p-1)\pm1]/3$ is
an integer. Thus, the
frequency of the first interband transition is given by
\begin{equation} \label{freq}
f^{(1)}_p \simeq \frac{\upsilon _F}{\mu_p\pi R_p}\,,\quad\mu_p=
\left\{
\begin{array}{ll}
1\,, & {\rm metallic\,\, shell}\\
3\,, & {\rm semiconducting\, \,shell}
\end{array}\right.\,.
\end{equation}
The foregoing formula has been confirmed by direct numerical simulation. \cite{Dyachkov}

The frequency regime below the interband-transition regime for an MWCNT is
dictated by the condition
\begin{equation}
f < f_e\,,
\end{equation}
where
\begin{equation}
f_e=\stackrel{\rm min} {p} \, \left\{f^{(1)}_p\right\}\,,
\end{equation}
 while the condition
\begin{equation}
f >  (2\pi\tau)^{-1}
\end{equation}
 establishes the frequency regime for long-range guided-wave
propagation. As a result, the frequency regime wherein long-range guided
waves can produce  geometric (antenna)
resonances is as follows:
\begin{equation}\label{eq14}
(2\pi\tau)^{-1} \lesssim f \lesssim f_e\,.
\end{equation}

At frequencies in the regime $f \gtrsim f_e$,
 the
interband transitions contribute strongly to the surface conductivity of
each shell such that the real and the imaginary parts of this quantity
are approximately equal in magnitude.   Guided
waves then get attenuated heavily. At frequencies in the regime $f
\lesssim 1/(2\pi\tau) $, attenuation of guided waves is caused by
fast electron relaxation in all shells.

\subsection {Guided waves in an infinitely long MWCNT}
\label{Num0}

Let us now examine guided-wave propagation at $f = 11.2$~THz in an
infinitely long MWCNT of type A consisting of $N = 13$ shells. The
relaxation time $\tau$ is taken to be vanishingly small.

The
shells numbered $p\in\left\{4,7,10, 13\right\}$ are metallic.
All interband transitions for these
shells occur at frequencies exceeding $31$~THz and therefore do not contribute to the effective
conductivity (per unit length) of the
MWCNT. The imaginary part of the surface conductivity of a metallic shell
is positive-valued and exceeds the real part in magnitude. Thus,
the necessary condition for the long-range propagation of guided waves is
satisfied.

The semiconducting shells labelled $p\in\left\{ 2,3,5, 6\right\}$
have negligible surface conductivity, and therefore do not
influence the scattering properties of the chosen MWCNT at
$11.2$~THz.

The surface conductivities of the semiconducting shells labelled
$p\in\left\{8, 9, 11, 12\right\}$ are dictated mainly by the
corresponding first interband transitions occurring at $f_p^{(1)}
\in\left\{ 36.5, 32.5, 26.7 , 24.4\right\}$~THz. At  $f =
11.2$~THz. the imaginary parts of the surface conductivities of
these shells are negative and several times smaller than the
surface conductivities of adjacent metallic shells. Therefore, the
surface current densities in these semiconducting shells are
smaller and oppositely directed with respect to their counterparts
in adjacent metallic shells. The real parts of the surface
conductivities  of these semiconducting shells are high enough to
cause large ohmic losses.

We considered only the two eigenmodes of guided-wave propagation
in the chosen MWCNT with the smallest retardation, labeled as
$GW1$ and $GW2$. They correspond to the two roots of the
dispersion equation (\ref{eq12}), identified as $h_1 $ and $h_2 $,
with the smallest real parts: ${\rm Re}(h_2 )
> {\rm Re}(h_1 )$. Of all eigenmodes, these two
  will mostly influence the scattering properties of the
finite-length MWCNT, as discussed in Sec.~\ref{num1}.

\begin{figure}[!htb]
\begin{center}
\includegraphics[width=3.4in]{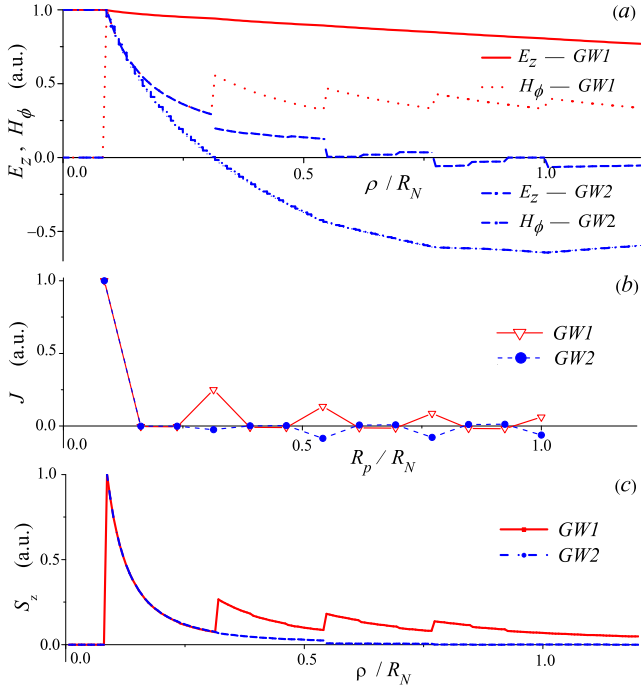}
\end{center}
\caption{The radial dependencies of (a) $E_z $ and   $H_\phi$, (b)
$J_p$ on the surfaces of the shells, and (c) the axial component
of the time-averaged Poynting vector,  for guided waves $GW1$ and
$GW2$, in an MWCNT of type A with $N = 13$ shells at $f = 11.2$
THz, in the limit  $\tau \rightarrow \infty$. The discrete points
in (b), corresponding to different shells,  are joined together
only to aid the eye.  } \label{fig2}
\end{figure}

The radial dependences of  $E_z$ and $H_\phi$ inside the MWCNT of
type A for guided waves $GW1$ and $GW2$ are shown in
Fig.~\ref{fig2}(a). Clearly, the axial component of the electric
field  is distributed over the entire cross section of the MWCNT.
The azimuthal component of the magnetic field is discontinuous
across  each metallic shell, in accordance with the boundary
condition (\ref{boundary3}). The degree of discontinuity decreases
as the shell number $p$ increases, in compliance with the
generally decreasing magnitudes of $J_p$ in Fig.~\ref{fig2}(b).
Outside the MWCNT, the radial distribution of the electric and
magnetic fields is governed by the argument of the modified Bessel
function $K_0 (\sqrt {h^2 - k^2 } \rho )$, whereby we conclude
that the electromagnetic field is highly localized to the MWCNT.

The axial surface current density $J_p$, $p\in\left[1,N\right]$,
is shown in Fig. \ref{fig2}(b)
for guided waves $GW1$ and $GW2$. In opposition to the radial distributions of
the axial electric
field, the magnitude of the current density is maximal on the  innermost
shell, and then strongly decreases with the increase of the shell
number $p$. This behavior is in agreement with the $R^{-1}$-dependence of the
surface conductivity of a metallic SWCNT of radius $R$.  \cite{Slepyan99}

Figure~\ref{fig2}(c) shows the radial distribution of the axial
component of the time-averaged Poynting vector
\begin{equation} \label{poynting}
S_z = \frac{c}{8\pi}\, {\rm Re}\left(\frac{h}{k}\right)\, \vert H_\phi \vert ^2
\end{equation}
for  $GW1$ and $GW2$, inside the MWCNT and in the
vicinity of its outermost shell. The
energy-flux density for $GW1$ is maximum near the surface of
the innermost metallic shell and then slightly varies about some mean
value between the 4$^{th}$ and the 13$^{th}$ shells. The energy-flux
density for $GW2$ is mostly concentrated between the two innermost
metallic shells and then decreases very rapidly as $\rho\to R_N$.
Clearly then, the two innermost metallic shells are dominant contributors to
 the retardation of both $GW1$ and $GW2$.

\begin{figure}[!htb]
\begin{center}
\includegraphics[width=3.4in]{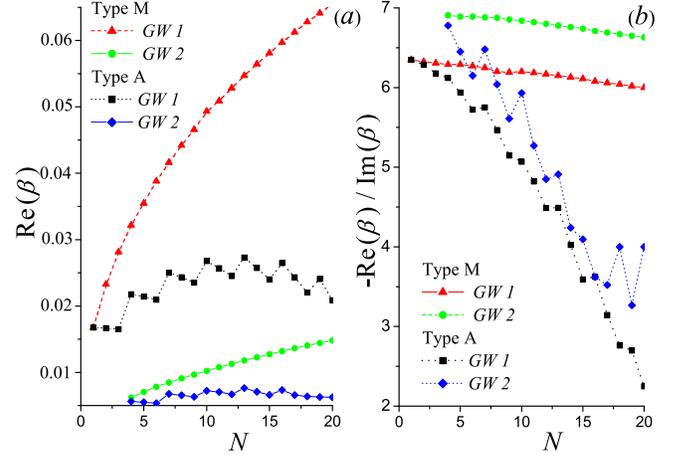}
\end{center}
\caption{Dependences of (a) ${\rm Re}(\beta)$  and (b) $ - {\rm
Im}(\beta ) / {\rm Re}(\beta )$  on $N$ for guided waves $GW1$ and
$GW2$ in MWCNTs of type A and M, when   $f = 11.2$~THz and $\tau =
10^{ - 13}$ s. Discrete points are joined by lines to aid the eye.
} \label{fig3}
\end{figure}

The dependences of the real part of the slow-wave coefficient $\beta $
and the ratio $ -{\rm Re}(\beta )  / {\rm Im}(\beta ) =  {\rm Re}(h)/
{\rm Im}(h)$ of the guided waves $GW1$ and $GW2$ on the number $N$
of shells in MWCNTs of types A and M are shown in Fig.~\ref{fig3}.
The retardation coefficient ${\rm Re}(\beta )$ for MWCNTs of type M
is higher than for MWCNTs of type A of comparable $R_N$, which is a significant observation
in relation to the geometric resonances of finite-length MWCNTs (Sec.~\ref{num1}).

For MWCNTs of type
M, ${\rm Re}(\beta)$  increases as $N$ does. As $N$ increases,
 the additional outmost shell has a lower influence on ${\rm Re}(\beta)$.
 The increase of ${\rm Re}(\beta)$  with increasing $N$
occurs until $R_N$ exceeds $\upsilon_F/\pi f$, per conditions
(\ref{eq14}) and (\ref{freq}).

Figure~\ref{fig3}(a) shows that, for guided waves $GW1$ and $GW2$ in MWCNTs of
type A, ${\rm Re}(\beta)$
\begin{itemize}
\item[(i)] increases with the addition of a
metallic shell, but
\item[(ii)] decreases with the addition of a semiconducting shell.
\end{itemize}
For an MWCNT  with small $N$, ${\rm Re}(\beta )$ is thus determined
mostly by the metallic shells. In contrast, for MW\-CNTs with large
$N$, both  ${\rm Re}(\beta )$ and $ -{\rm Re}(\beta ) / {\rm
Im}(\beta )$ are strongly affected by the semiconducting shells of
large radius. This occurs because the semiconducting shells labelled
$p\in\left\{ 11, 12, 14, 15, 17, 18, 20\right\}$ have their first
interband transitions near the chosen frequency of $11.2$~THz, and,
consequently, the real and imaginary parts of their surface conductivities
are large in magnitude. An increase in the number of such shells
greatly diminishes the parameters ${\rm Re}(\beta)$ and $ -{\rm
Re}(\beta )  / {\rm Im}(\beta )$ of $GW1$ and $GW2$ in the MWCNTs of type
A. Thus, \textit{interband transitions suppress guided-wave
propagation in MWCNTs}.

\subsection{Scattering properties of a finite-length MWCNT}
\label{num1}

Let us now move on to the scattering properties of finite-length MWCNTs in the terahertz and the
far-infrared frequency regimes. Here we focus only on the case when the
incident electric field is parallel to the $z$ axis, thereby permitting us
 to investigate electromagnetic effects caused
by the axial surface conductivities of the shells.

\begin{figure}[!htb]
\begin{center}
\includegraphics[width=3.4in]{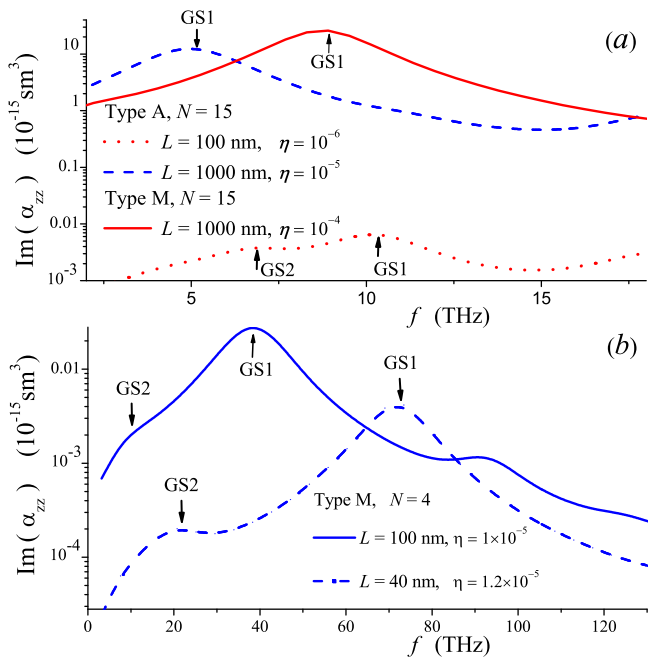}
\end{center}
\caption{Frequency-dependence of ${\rm Im}(\alpha_{zz} )$ of
MWCNTs of type A and M for different lengths $L$ and number of
shells $N$. The labels $GS1$ and $GS2$  denote the first geometric
resonance --- $s = 1$ in  (\ref{eq15}) --- of the guided waves
$GW1$ and $GW2$, respectively.   (a) $\tau = 10^{ - 13}$~s, (b)
$\tau = 2 \times 10^{ - 14}$~s.} \label{fig4}
\end{figure}

Figure~\ref{fig4}(a) illustrates the frequency dependence of the
imaginary part of the polarizability scalar $\alpha_{zz} $ in the
long-wavelength regime ($kL \ll 1$) for different lengths $L$ and
shell numbers $N$ in MWCNTs of type A and M. The labels $GS1$ and $GS2$ in
this figure denote the first geometric resonance of the guided
waves $GW1$ and $GW2$, respectively.  The geometric resonances
occur at frequencies \cite{Slepyan06,Shuba07}
\begin{equation}\label{eq15}
    f_s = s \left(\frac{c}{2L}\right) {\rm Re}(\beta)\, , \quad s\in\left\{1,3,5,...\right\}\,.
\end{equation}
As we concluded from Fig.~\ref{fig3}(a),
the retardation coefficient ${\rm Re}(\beta )$ for MWCNTs of type M
is higher than for MW\-CNTs of type A of comparable $R_N$. Therefore the first geometric resonance
($s=1$) for
the MWCNTs of type M appears in Fig~\ref{fig4}(a) at a higher frequency than for MWCNTs of
type A, both of the same length $L$.

The resonance frequencies for   $GW1$ and $GW2$ depend on $L$
nonlinearly. As an example, in Fig.~\ref{fig4}(a)  the  resonance
frequency of the MWCNT of type A increases by a factor of about $2$
(from $5$ to $\sim10$~THz) while the  length $L$ decreases by a
factor of $10$   (from $1000$ to $100$~nm). That is reflected in
(\ref{eq15}) by the strong dependence of  ${\rm Re}(\beta)$ on $f$,
when  $f$ is close to $f_e$. Accordingly, the dependence of the frequency of the
geometric resonance on $L$ is not explicit.

The first resonance of the MWCNT of type A and length $L=100$ nm at
$\sim10$-THz frequency is not strong, because condition (\ref{eq14}) is not
satisfied for this MWCNT, and the guided wave is strongly attenuated.
We conclude that the resonance of an MWCNT antenna of type A or M near a given
frequency are the most pronounced, if both (\ref{eq14}) and (\ref{eq15})
are satisfied.

Figure~\ref{fig4}(a) contains the value of the antenna efficiency
$\eta$ at the first geometric resonance for all MWCNTs considered.
This antenna efficiency depends both on the type and the length of
the MWCNT. The antenna efficiency of an MWCNT is several time larger
than that of an SWCNT of the same length. Our calculations make us conclude
 that the restrictions
on the MWCNT dimensions given by (\ref{eq14}) and (\ref{eq15}) do
not permit an increase in $\eta$ at the first geometric resonance.
In contrast, the antenna efficiency of an almost circular bundle of closely packed SWCNTs can be
increased up to unity, by increasing of number of metallic SWCNTs in
the bundle. \cite{Shuba07}

When conduction in an MWCNT is very diffuse,
\cite{Lee} the relaxation time $\tau$ is close to that for graphite
($ 2 \times 10^{ - 14}$~s). Then the conditions (\ref{eq14}) and
(\ref{eq15}) can be fulfilled only for MWCNTs of type M, and that
too with $R_N \lesssim 2.5$ nm and $L \lesssim 200$ nm.
Figure~\ref{fig4}(b) shows the frequency dependence of ${\rm
Im}(\alpha_{zz} )$ of an MWCNT of type M with $N = 4$ for $\tau = 2
\times 10^{ - 14}$ s. The first geometric resonances of MWCNTs of
lengths $L = 100$ nm and $L = 40$ nm appear in the far-infrared
($f = 38$ THz) and the mid-infrared ($f = 72$ THz) regimes,
respectively. Thus, antenna resonances are pronounced and can be
experimentally observed for short MWCNTs with several shells only in the
far-infrared and mid-infrared regimes.

\subsection{MWCNT properties in the interband-transitions regime}
 \label{num2}

Whereas Secs.~\ref{Num0} and \ref{num1} address the frequency
regime $f\in\left(1/2\pi\tau,f_e\right)$ wherein interband transitions
are not possible, we now proceed to the interband-transitions regime of
Sec.~\ref{low-coupling}, wherein the electromagnetic response properties of MWCNTs are dominated
by the interband transitions.
This regime, delineated by the condition
$f>f_e$, can very well lie in the visible part of the electromagnetic spectrum.

Wang {\it et al.} have published the only known
report on the experimental observation of antenna resonances of MWCNTs
in the visible regime.\cite{Wang_04_APL}
In their experiment,
interference patterns of light scattered from an array of finite-length
MWCNTs with $R_N \approx 25$ nm were observed. Wang {\it et al.} attributed the
 observed phenomenon to
a \textit{length-matching} antenna effect, which requires the
spatial variations of the induced current density to satisfy the
edge conditions; furthermore, they assumed that a thick MWCNT has
the same scattering response in the far zone as a perfect
conducting cylinder with the same cross-sectional radius and
length. However, Hao and Hanson \cite{Hanson07} subsequently
showed that the model of an MWCNT as a single cylindrical shell
with typical surface conductivity  given, e.g., by  (\ref{eq4}),
can not describe the experimental observation recorded in
Ref.~\onlinecite{Wang_04_APL}. The reason is the low surface
conductivity of the MWCNT shells, as given by (\ref{eq4}). In the
model of Ref.~\onlinecite{Hanson07}, however, the intershell
electromagnetic coupling in the MWCNT was ignored, which is
inappropriate if $R_N \gtrsim 20$~nm, as we pointed out in
Sec.~\ref{low-coupling}.

The lacuna in the model of Ref.~\onlinecite{Hanson07} can be removed by
implementing the integral-equation technique of Sec.~\ref{inteq}. Therefore, we
decided to examine the
scattering properties of finite-length MWCNTs of radius $R_N \simeq 25$
nm and $R_N \simeq 50$ nm in the visible regime, taking the intershell electromagnetic coupling
into account (but ignoring the intershell electron tunneling).
As we discuss later in this subsection, our analysis confirms the absence of
the {length-matching} antenna effect in the
 near-infrared and the visible regimes.

\begin{figure}[!htb]
\begin{center}
\includegraphics[width=3.4in]{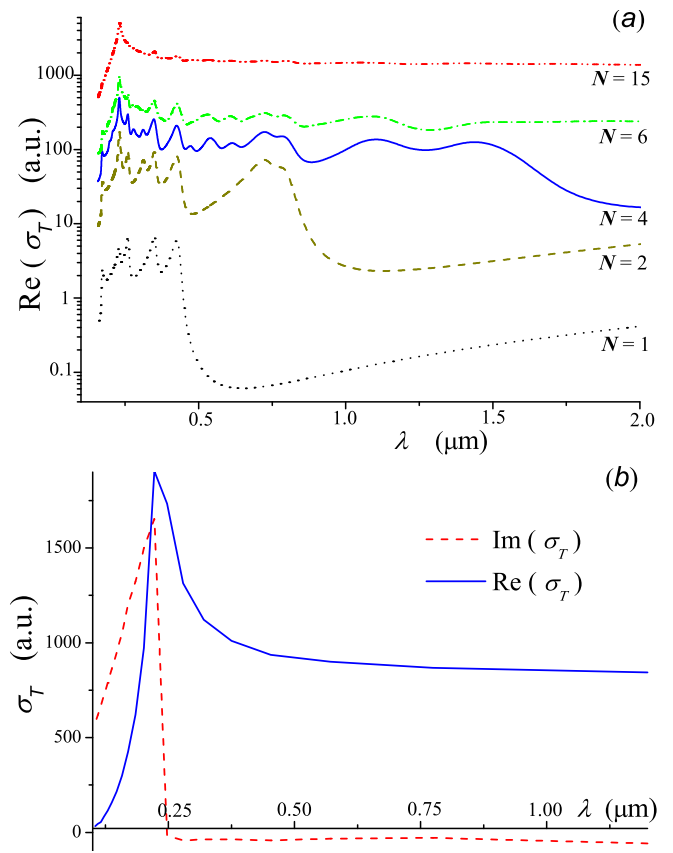}
\end{center}
\caption{(a) Real part of the effective conductivity per unit
length of an MWCNT of type M in the near-infrared and the visible
regimes. The number $N$ is variable, and $\tau = 2 \times 10^{ -
14}$~s. Note that $R_{15} =5.284$~nm. (b) Real and imaginary parts
of $\sigma_T$ of an MWCNT of type M with $N = 29$ shells in the
near-infrared and the visible regimes. } \label{fig5}
\end{figure}

Let us first analyze the  effective conductivity per unit length
of an electrically thin MWCNT, as estimated by  (\ref{sigm}).
Figure \ref{fig5} (a) presents the real part of $\sigma _T $ as a
function of $N$ for MWCNTs
 in the near-infrared and the visible
regimes. This figure shows that, for SWCNTs ($N=1$)
and MWCNTs with $N \in\left[ 2, 4\right]$,  $\sigma _T $ has strong
resonances. The resonances weaken as $N$ increases and practically disappear
for  $N > 10$. This trend can be explained in the following
way. In the near-infrared and the visible regimes, the surface
conductivity $\sigma _p$ of the $p$-{th} shell, obtained from
(\ref{eq4}), has many resonances corresponding to
van Hove singularities. As the resonances of different shells overlap,
 the weighted summation of the surface conductivities
of all shells in   (\ref{sigm}) ensures that  $\sigma _T $ does
not evince resonant behavior in the near-infrared and the visible
regimes.

The conductivity per unit length of an isolated shell of radius
$R_p \gtrsim 5$ nm also has a large number of closely spaced
resonances so that, instead of discrete lines, a band appears in
its spectrum. As the surface conductivities of all MWCNT shells
have a plasmon resonance \cite{Shuy} in the ultraviolet regime at
the free-space wavelength $\lambda_{pl} = (c \pi \hbar)/\gamma
_0$, $\sigma _T$ for any MWCNT also has a resonance at this
wavelength.

 The real and imaginary parts of the effective per-unit-length conductivity $\sigma _T $
for an MWCNT of type M with $N=29$ shells (i.e., $R_N= 10.22$~nm)
are presented in Fig.~\ref{fig5}(b). As  this figure demonstrates,
the real part of $\sigma _T $ increases as the frequency increases
and has a maximum near $f = 1304$ THz (i.e., $\lambda = 230$ nm), which is not a geometric-resonance
frequency but, instead, is a
plasmon-resonance frequency. \cite{Shuy}  In the near-infrared and the visible regimes,
the condition ${\rm Re}(\sigma _T ) \gg {\rm Im}(\sigma _T )$ holds;
therefore, the chosen
 MWCNT cannot support surface-wave propagation, which occurs for a
metallic wire with   ${\rm Re}(\sigma ) \ll {\rm Im}(\sigma )$
according to Ref.~\onlinecite{Pitarke}. We also found that the
electric field exciting a particular shell differs very slightly
from the electric field incident on the MWCNT, when $R_N\lesssim
20$~nm   and the
frequency lies in either the near-infrared or the visible regimes,
thereby implying that the electromagnetic coupling between the
shells is slight.   The frequency dependence of the scattering power $P_t$
for such an MWCNT is the same as of $|\sigma_T(\omega)|^2$, according to
(\ref{eq21c}).

Let us now carry on to electrically thick MWCNTs (with $R_N\simeq
25$~nm or $50$~nm). For calculation of the electric current
densities in their shells in the near-infrared and the visible
regimes we used (\ref{eq18a}), with $E_{z}^{(0)}(\rho,z)$ assumed
to be independent of $\rho\in\left[0,R_N\right]$.
 Such an approximation is sufficient to ascertain
whether  geometric resonances of azimuthally symmetrical modes exist
in thick MWCNTs or not.  Of course, $\sigma_T$ cannot be defined for
electrically thick MWCNTs.

\begin{figure}[!htb]
\begin{center}
\includegraphics[width=3.4in]{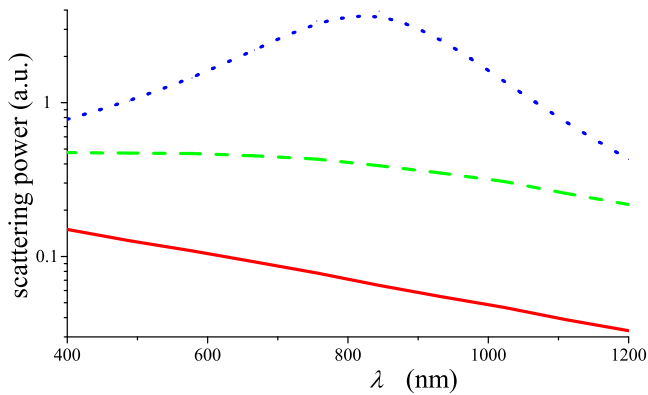}
\end{center}
\caption{Scattered power $P_r$ versus free-space wavelength
$\lambda$ for MWCNTs with $N = 70$ (solid line) and $N = 140$
shells (dashed line) and length $L = 350$~nm. The scattered power
for a perfectly conducting rod (dotted line) of cross-sectional
radius 25 nm and length 350 nm was calculated by solving
(\ref{eq18a}) with $N = 1$ and $\sigma_1 \rightarrow \infty $.}
\label{fig6}
\end{figure}

\begin{figure}[!htb]
\begin{center}
\includegraphics[width=3.4in]{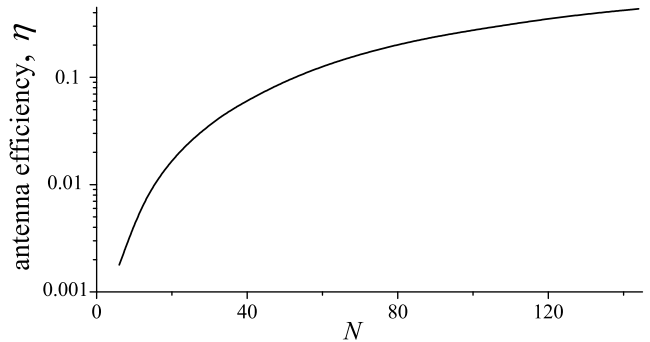}
\end{center}
\caption{Dependence of the antenna efficiency $\eta$ on the number
$N$ of shells in an MWCNT of type M and length $L = 350$ nm at
frequency $f = 600 $~THz (visible regime).} \label{fig7}
\end{figure}

In the near-infrared and visible regimes, the surface conductivity of a shell of large radius
 ($R\gtrsim 5$ nm), in accordance with
(\ref{eq4}), does not depend on that whether shell is metallic or
semiconducting.
Therefore, though the plots in Figs. \ref{fig6}-\ref{fig8}
were made for thick MWCNTs of type M, all results presented
therein are qualitatively true for thick MWCNTs of type A also.

In order to compare the electromagnetic responses of the chosen
MWCNTs and a perfectly conducting rod in the near-infrared and the
visible regimes, we need to calculate the scattered power $P_r$
when the incident electric field is oriented parallel to the $z$
axis. When we calculated the scattered power $P_r$ for MWCNTs of
type M, length $L = 350$~nm,  and $N=70$ ($R_N=24.66$~nm) or
$N=140$ ($R_N=49.32$~nm), resonances did not show up for
$f\in\left[250,750\right]$~THz (i.e., $\lambda\in[400,1200]$~nm)
in Fig. \ref{fig6}. For comparison, the scattered power for a
perfectly conducting rod of cross-sectional radius 25~nm and
length 350~nm is also shown  in Fig. \ref{fig6}. This nanorod antenna,
in contrast to the MWCNTs, has a set of resonances determined by
the condition
\begin{equation}\label{eq17}
    L = \kappa s\lambda / 2\,,\quad s\in\left\{1,2,3,....\right\}\,,
\end{equation}
where $\kappa$ is a correction factor that
slightly exceeds unity  and is a function of the ratio of the length
and the wavelength as well as of the ratio of cross-sectional radius and the length.
 \cite{balanis} For the chosen nanorod antenna, the first resonance ($s = 1$) is characterized by $\kappa =
1.15$ and appears at $f=370$~THz (i.e., $\lambda = 810)$~nm, which
is confirmed by the dotted line Fig.~\ref{fig6}.

The absence of antenna (geometric) resonances of the chosen MWCNTs in the visible regime
can be explained by the strong dissipation of electromagnetic energy in
MWCNT shells and the small electromagnetic coupling between the shells, as discussed in
Sec.~\ref{low-coupling}.
Therefore, even an MWCNT with $R_N \simeq 50$ nm and
comprising 140 shells can not support guided-wave
propagation  and, consequently, can not display the
length-matching antenna effect. The same
conclusion is also true in the visible regime for SWCNTs \cite{Slepyan06} and
planar arrays of finite-length SWCNTs. \cite{Hanson07} Let us remark that a
hypothetic multishell conductive structure with $N = 70$ and $R_N \simeq
25 $ nm can have an antenna resonance corresponding to $s=1$ in (\ref{eq17}),
provided the surface conductivity of every shell is five times that given
 by (\ref{eq4}); but such a structure has not been practically realized as
 yet.

Our approach, which is enhanced with respect to that of Hao and Hansen, \cite{Hanson07}
can not explain the MWCNT antenna resonance
reported by Wang {\it et al.} \cite{Wang_04_APL} Additional experiments are necessary
to resolve the contradiction between interpretation of measurements
proposed by Wang \emph{et al.}  and  the theoretical
model of MWCNTs developed here and based on   Ref.~\onlinecite{Slepyan99}.

The dependence of the antenna efficiency $\eta$ in the visible
regime on the number $N$ of shells in the MWCNT  is illustrated in
Fig.~\ref{fig7}. The antenna efficiency increases with the number
of shells and tends to unity for thick MWCNTs; indeed,  $f =
600$~THz, we calculated $\eta = 0.17$ for $N=70$, but $\eta=0.44$
for $N=140$.

\subsection{Surface-plasmon-wave propagation in an MWCNT with a gold core}
 \label{num3}

These days, thin metallic (gold, silver, and
aluminium)  wires of  finite length  are considered to be promising for application
as optical nanoantennas. \cite{Novotny, Novotny07} However, the fabrication of
long, high-quality, thin, single-crystal wires of cross-sectional radius less
than 5 nm and with perfect cylindrical form (i.e., without breaks, bends,
deformations, etc.) is a difficult technological problem. Recently,
CNTs have been used as templates in order to promote the
formation of high-quality single-crystal wires coated by
perfect graphene cylinders. \cite{Elias} This is an exciting possibility for a composite nanoantenna
comprising a solid metallic core covered by
concentric carbon shells. Clearly, such a structure is neither a metallic cylinder nor an
isolated MWCNT.
\begin{figure}[!htb]
\begin{center}
\includegraphics[width=3.4in]{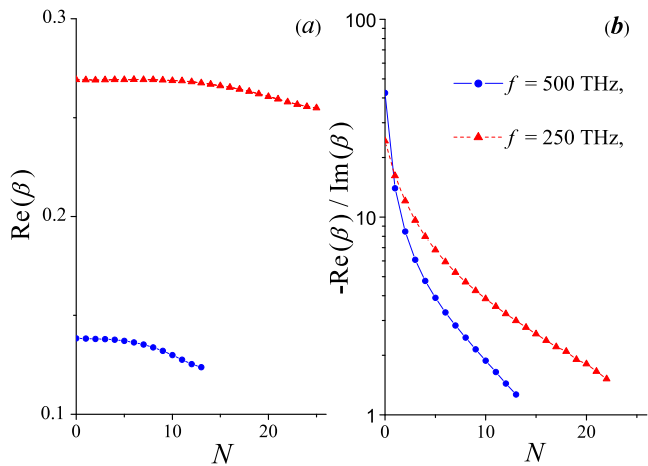}
\end{center}
\caption{Dependences of (a) ${\rm Re}(\beta)$  and (b) the ratio $
- {\rm Re}(\beta ) /{\rm Im}(\beta )$  on the number $N$ of carbon
shells for a surface-plasmon wave in an MWCNT with a gold core.
The MWCNT is of type M, and its $p$-th shell has a $(90+9p,0)$
zigzag configuration and radius $R_p =
 \sqrt{3}(90+9p)b/(2\pi)$,  $p\in\left[1,N\right]$. The gold core has a
 cross-sectional radius $R_0=3.5$~nm.
 Data for an isolated metal wire of cross-sectional radius $R_0$ is the same
 as for $N=0$.
Discrete points are joined by lines to aid the eye. } \label{fig8}
\end{figure}

Surface-plasmon waves in the infrared and the visible regimes can propagate along a metal
wire. \cite{Pitarke} Surely, guided-wave propagation would be affected if the metallic wire
were to be enclosed in an ensemble of
concentric carbon shells. The theoretical approach of Sec.~\ref{guided} can
be applied to study the phenomenon of surface-plasmon waves as follows.

Suppose that the metal core is of cross-sectional  radius $R_0 < R_1$, the radius of the innermost shell
of an MWCNT. If  $R_0$ is much less than
 the skin depth of the bulk metal and $kR_0\ll 1$,
 then the metal core   can be modeled as a solid cylinder with
 surface conductivity $\sigma_0 = \sigma_{met}R_0/2$, \cite{Hanson061} where $\sigma_{met}$ is the bulk
 volumetric conductivity
 of the metal. Also,  $R_0$ has to
be higher than a critical value $R_{cr}$,
  which corresponds to the crystalline-noncrystalline
 transition in metals and separates the quasi-bulk behavior ($R_0 > R_{cr}$) from the quasi-molecular
behavior ($R_0 < R_{cr}$) of a nanowire.
Then,
(\ref{boundary3})--(\ref{current}) also hold at $\rho = R_0$ with
$\sigma_p \rightarrow
\sigma_0$, and the procedures of Sec.~\ref{guided} are applicable.

We considered the propagation of surface-plasmon waves in an MWCNT
with a gold core. The bulk volumetric conductivity $\sigma_{met}$ of gold was
taken to   follow the Drude model with parameters given in
Ref.~\onlinecite{Johnson}. The skin depth of gold in the visible
regime is higher than $30$~nm. The value $R_{cr} = 1.5$~nm for
gold was obtained with first-principles calculations in
Ref.~\onlinecite{Wang01}. So we assumed that $1.5$ nm $ < R_0 \ll
30$~nm.

Figure~\ref{fig8} presents the dependences of ${\rm Re}(\beta)$
and $ - {\rm Re}(\beta ) /{\rm Im}(\beta ) $ of a surface-plasmon
wave on the number $N$ of shells
 in MWCNT of type M with a gold core of cross-sectional radius $R_0=
 3.5$~nm.  The $p$-th shell of the MWCNT
 has a $(90+9p,0)$ zigzag configuration and radius $R_p =
 \sqrt{3}(90+9p)b/(2\pi)$,  $p\in\left[1,N\right]$.
 In the near-infrared ($f=250$ THz) and the visible ($f=500$ THz)
regimes, Fig.~\ref{fig8}(a) shows that the retardation coefficient
${\rm Re}(\beta)$ depends only slightly on $N$. According to
Fig.~\ref{fig8}(b), the value of $ - {\rm Re}(\beta ) /{\rm
Im}(\beta ) $ is less than for an isolated metal wire ($N=0$)  and significantly decreases as $N$ increases.

These trends can be explained in the following manner. The guided
wave propagates partly inside the MWCNT and partly inside the gold
core.  The dissipation in the MWCNT shells is high because the real and the
imaginary parts of the surface conductivities of the
shells in the interband regime are of comparable magnitudes.
 This dissipation increases with
frequency, in accordance with the frequency dependence of the
effective per-unit-length conductivity of the MWCNT presented in
Fig.~\ref{fig5}(b).
 Because of the weak electromagnetic coupling of CNT
shells, the shells strongly absorb electromagnetic energy
independently on each other --- which explains the strong
dependence of $ - {\rm Re}(\beta ) /{\rm Im}(\beta ) $ on $N$.
Furthermore, the weak coupling changes the radial electromagnetic
field distribution of the surface-plasmon wave only slightly as
$N$ increases, which explains the weak dependence of ${\rm
Re}(\beta)$ on $N$.

The decrease of the magnitude of $ - {\rm Re}(\beta ) /{\rm Im}(\beta ) $
with increasing $N$ increasing   implies the enhancement of $P_t$, the
power lost to ohmic dissipation,  and consequently the
decreasing of the antenna efficiency  $\eta$ of an MWCNT with a metal core.
However, if $N$ is not too large ($N = 2$ or
$3$) some worsening of antenna properties may be justified by other
advantages that the metal-core MWCNT may confer.

\section{Discussion}
\label{Discus}

Thus, an MWCNT can operate as an antenna in two different regimes.
The first  is the
\emph{Drude-conductivity regime}, wherein the motion of conduction-band  electrons
is responsible for radiation properties. This regime has a distinct analogy
with a classical radio-frequency wire antenna, as is clear from Sec.~\ref{low-coupling}
and \ref{num1}.
 The
second regime is the \emph{interband-transitions regime}, wherein
quantum transitions of electrons between
different energy states occur. This regime was considered
in Sec.~\ref{num2} and does not have a classical analogy. The frequency $f_e$ separates the Drude-conductivity regime
$f\in\left(1/2\pi\tau,f_e\right)$
from the interband-transitions regime $f>f_e$.

Guided-wave propagation and geometric resonances of the guided waves
are typical for  macroscopic wire antennas. The guided waves have a
quasi-transverse-electromagnetic structure and are characterized by
low retardation and low attenuation. \cite{balanis} The existence of
guided (surface) waves and geometric resonances also is typical for
nanowire antennas in the Drude-conductivity regime.
\cite{Novotny07,Slepyan99,Slepyan06} But the guided wave has strong
retardation and high attenuation, so that the frequency of a geometric
resonance is not connected to the free-space wavelength but to a
shorter effective wavelength that depends on the material
properties. \cite{Novotny} This general rule is also true for MWCNT
antennas:  Fig. \ref{fig3} shows
that guided waves have strong retardation and high
attenuation, and Fig.~\ref{fig4} presents geometric resonances at $\lambda
\ll  L$ demonstrating thereby the effective wavelength to be shorter than
free-space wavelength.

Calculated data presented in Fig.~\ref{fig3}(a) indicate that the
retardation coefficient increases when the number of shells
increases. Furthermore, the retardation coefficient is the highest
for the $GW1$ mode in MWCNTs of type M. That implies that a MWCNT
with only metallic shells and operating in the $GW1$ mode offers
attractive prospects for high antenna efficiencies in the
terahertz regime.

The frequency $f_e$, separating the Drude-conductivity regime from
the interband-transitions regime, depends on the detailed
electronic and geometric attributes of the MWCNT. According to
(\ref{freq}), $f_e$ decreases as $R_N$ increases. As examples, for
an MWCNT of type A, (i) $R_N = 10$ nm and $f_e  = 9.3$~THz when
$N=32$, but  (ii) $R_N = 1.9$ nm and $f_e  = 48$~THz when $N=6$.
In the interband-transitions regime $ f > f_e $, guided-wave
propagation and geometric resonances are absent for both SWCNTs
and MWCNTs, which cardinally distinguishes this regime from  the
Drude-conductivity regime.

As shown experimentally,\cite{chin} the absorption and the
scattering characteristics of an electrically thick MWCNT in the
regime of optical transitions depend only slightly on the
frequency. This conclusion, also borne out by the data in
Fig.~\ref{fig6}, may be seem to be unexpected. In fact optical
transitions are resonance processes and the surface conductivities
of SWCNTs and MWCNT shells have resonances corresponding to the
van Hove singularities. But we found that the antenna
parameters of electrically thin MWCNTs are determined by the
effective parameter $\sigma _T $ defined in (\ref{sigm}). The
overlapping of a large number of resonances of the surface
conductivities of the different shells leads to a smooth frequency
dependence of $\sigma _T $. Such an effect is analogous to
inhomogeneous broadening in an ensemble of all different harmonic
oscillators.

Thus, within the framework of our model, it is impossible to
interpret the observation scattering resonances
of random arrays of MWCNTs with average outermost radius $25$~nm
reported in Ref.~\onlinecite{Wang_04_APL}. The same conclusion emerged from
the model of Hao and Hansen.
\cite{Hanson07} Further experiments are required.

Antennas are  objects that transform a near-field into a far-field
and vice versa. The morphology of the near field of a nanoantenna
possess a nanoscale, and is therefore determined by quantum size
effects. Therefore the coupling of a nanoantenna with its near field
is stronger than that of a macroscopic antenna with its near field;
as a result, the transformation of the near field to the far field
by a nanoantenna is more difficult, and the antenna efficiency of a
thin-nanowire antenna is low. This property is inherent to different
types of nanoantennas: SWCNTs \cite{Hanson05,Burke06}, bundle of
SWCNTs \cite{Shuba07}, metallic nanorods \cite{Hanson08}, and MWCNTs
as in Fig.~\ref{fig4}. The antenna efficiency can be enhanced by
increasing the number of shells in an MWCNT (Fig. \ref{fig7}), the
number of SWCNTs in a bundle (Fig. 7 in Ref.~\onlinecite{Shuba07}),
and the cross-sectional radius of a metallic rod antenna (Fig. 1 in
Ref.~ \onlinecite{Hanson08}). We can thus conclude that the small
value of antenna efficiency is the fundamental physical
characteristic of nanoantennas. Nevertheless, its antenna efficiency
is tunable over a wide range.

In order to achieve antenna efficiency close to unity, it is
necessary to strongly suppress the influence of quantum size effects
by ensuring that the cross-sectional radius is high. Note that, since
quantum size effects are not pronounced in gold nanowires of
cross-sectional radius of several tens of nanometers, such
nanowire-based antennas are expected to possess properties analogous
to macroscopic antennas.

The electromagnetic properties of MWCNTs have only a slight
frequency-dependence in the interband-transitions regime, per
Fig.~\ref{fig6}. Thus, and MWCNT can be considered to be a
nanoantenna
 with sufficiently high $\eta $ ($
\approx 0.1$ in Fig.~\ref{fig7}) and a wide operating-frequency range in the visible regime. Such
nanoantennas have properties similar to those of electrically small but macroscopic
antennas in microwave regimes (e.g., short
non-resonant dipoles). \cite{balanis}

\section{Concluding Remarks}
\label{concl} To conclude, we modeled the shells of an MWCNT as
impedance sheets with axially directed surface conductivity,
ignored intershell tunneling of electrons but incorporated
intershell coupling, in an integral-equation approach. Calculated
data indicate that in a low-frequency regime called the
Drude-conductivity regime, wherein optical interband transitions
do not occur,  guided waves can propagate with low attenuation in
an MWCNT which has  metallic shells. In the same frequency regime,
the axial polarizability of a finite-length MWCNT has a resonant
behavior due to the antenna-length matching effect. However, the
shells with surface conductivity due to interband transitions
suppress guided-wave propagation. Due of the high dissipation in
such shells, MWCNTs with outermost radius $\approx 25$ nm can not
possess resonant properties in the visible regime. Analysis of
surface-plasmon-wave propagation in a MWCNT with a gold core shows
that, in the near-infrared and the visible   regimes, the shells
behave effectively as  lossy dielectric materials and suppress
surface-wave propagation along the gold core.

The following conclusions regarding the operation of
MWCNTs as nanoantennas emerged from our work:
\begin{itemize}
\item[(i)] The antenna efficiency $\eta$ of an MWCNT exceeds that
of an SWCNT but is less than that of an almost circular bundle of
closely packed, metallic SWCNTs, provided that all three objects
are of roughly the same outermost radius. Therefore, SWCNT-bundles
are the most promising candidates for terahertz and mid-infrared
antennas.

\item[(ii)]An MWCNT with at least 4 shells is recommended for
application as a nanoantenna with a wide operating-frequency range
in the visible regime.

\item[(iii)] Filling the core of an MWCNT
with a metal makes the MWCNT attractive as a  nanoantenna,
provided that the number of shells does not exceed 3.
\end{itemize}

The model developed in this paper can
be applied for an MWCNT with achiral shells and negligibly small
intershell tunneling. Generalization of the  boundary
condition (\ref{boundary3})-(\ref{eq4}) is needed to take the chirality of shells
and intershell tunneling into account.

\acknowledgments

This research was partially supported by INTAS under projects
05-1000008-7801 and 06-1000013-9225, International Bureau BMBF
(Germany) under project BLR 08/001, and the Belarus Republican
Foundation for Fundamental Research and CNRS (France) under project
F07F-013. MVS thanks the World Federation
of Scientists for a fellowship. AL acknowledges the Charles Godfrey Binder Professorship Endowment at the
Pennsylvania State University for partial support.


\begin{thebibliography}{99}

\bibitem{Dressel_B01} M. S. Dresselhaus, G. Dresselhaus, and Ph. Avouris,   \textit{Carbon Nanotubes\/} (Springer, Berlin, Germany,  2001).

\bibitem{Reich_b04} S. Reich, C. Thomsen, and J. Maultzsch,  \textit{Carbon Nanotubes. Basic Concepts and Physical Properties\/} (Wiley-VCH, Berlin, Germany,
2004).

\bibitem{Health}
M. L. Schipper, N. Nakayama-Rachford, C. R. Davis, N. W. S. Kam, P. Chu,
Z. Liu, X. Sun, H. Dai,  and S. S. Gambhir, {Nature Nanotechnol.} \textbf{3}, 216 (2008).


\bibitem {Slepyan99} G. Ya. Slepyan, S. A. Maksimenko, A. Lakhtakia, O. Yevtushenko, and A. V. Gusakov,  Phys. Rev. B \textbf{60}, 17136 (1999).

\bibitem{Maksim00} S. A. Maksimenko and G. Ya. Slepyan,
%Electrodynamic
%properties of carbon nanotubes,
in \textit{Electromagnetic Fields in Unconventional Materials and
Structures\/} (O. N. Singh and A. Lakhtakia, eds.), (Wiley, New
York, NY, USA, 2000), pp. 217-255.

\bibitem{Maksim1}  S. A. Maksimenko and G. Ya. Slepyan,
Nanoelectromagnetics of low-dimensional structures, in
\textit{Nanometer Structures: Theory, Modeling, and Simulation\/}
(A. Lakhtakia, ed.), (SPIE Press, Bellingham, WA, USA, 2004), pp.
145-206.

\bibitem{Hagmann_05} M. J. Hagmann, IEEE Trans. Nanotechnol. \textbf{4},
289 (2005).

\bibitem{Rybczynski_07_APL} J. Rybczynski, K. Kempa, A. Herczynski, Y. Wang, M. J. Naughton,
Z. F. Ren, Z. P. Huang, D. Cai,  and M. Giersig, { Appl. Phys.
Lett.}~{\bf 90} 021104 (2007).


\bibitem{Raychowdhury_06} A. Raychowdhury and  K. Roy, IEEE Trans.
CAD Integrat. Circ. Syst.
\textbf{25}, 58 (2006).

\bibitem{Chiariello_07} A. G. Chiariello and G. Miano, COMPEL: Int. J. Comp. Math.
Electrical Electron. Engg. \textbf{26} 571  (2007).

\bibitem{Maffucci_08} A. Maffucci, G. Miano, and F. Villone, Int. J.
Circ. Theory Appl. \textbf{36}, 31 (2008).

\bibitem{Li_08} H. Li, W.-Y. Yin, K. Banerjee, and J.-F. Mao, IEEE Trans.
Electron Devices \textbf{55}, 1328 (2008).

\bibitem{Wang_04_APL} Y.~Wang, K.~Kempa, B.~Kimball, J.~B. Carlson,
G.~Benham, W.~Z. Li, T.~Kempa, J.~Rybczynski,   A.~Herczynski, and Z.~F. Ren,  Appl. Phys. Lett.~{\bf 85},  2607 (2004).

\bibitem {Hanson05} G. W. Hanson, {IEEE Trans. Antennas Propagat.} \textbf{53}, 3426 (2005).

\bibitem {Burke06} P. J. Burke, S. Li, and Z. Yu, IEEE Trans. Nanotechnol. \textbf{5}, 314
(2006).

\bibitem {Slepyan06} G. Ya. Slepyan, M. V. Shuba, S. A. Maksimenko,
and A. Lakhtakia, {Phys. Rev. B} \textbf{73}, 195416
(2006).

\bibitem{Kempa_07} K. Kempa, J. Rybczynski, Z. Huang, K. Gregorczyk, A. Vidan, B. Kimball, J. Carlson, G. Benham,
Y. Wang, A. Herczynski, and Z. F. Ren,  Adv. Mater. \textbf{19},
421 (2007).

\bibitem{Wang_08_IJIRMW}Y. Wang, Q. Wu, W. Shi, X. He, X. Sun,  and T. Gui, Int. J. Infrared
Millim. Waves  \textbf{29}, 35 (2008).

\bibitem{Hanson08} G. W. Hanson, {IEEE  Antennas Propagat. Mag.} in press, June (2008).

\bibitem {Maksimenko07} S. A. Maksimenko, G. Ya. Slepyan, A. M. Nemilentsau, and M. V.
Shuba, {Physica E} \textbf{40}, 2360  (2008).


\bibitem{Rutherglen} C. Rutherglen and P. Burke,
Nano Lett.  \textbf{7}, 3296 (2007).

\bibitem{Iensen} K. Jensen, I. Weldon, H. Garcia, and A. Zettl,
Nano Lett.  \textbf{7}, 3508 (2007); corrections: \textbf{8}. 374 (2008).

\bibitem{Misewich_03_Sci} J. A. Misewich, R. Martel, Ph. Avouris, J. C. Tsang, S. Heinze, and J. Tersoff,
%Electrically Induced Optical Emission from a Carbon Nanotube FET
Science  \textbf{300}, 783 (2003).


\bibitem{Chen_05_Sci} J. Chen, V. Perebeinos, M. Freitag, J. Tsang, Q. Fu, J. Liu, and Ph. Avouris,
%Bright Infrared Emission from Electrically Induced Excitons in Carbon Nanotubes.
Science \textbf{310}, 1171 (2005).

\bibitem{Kibis3} O. V. Kibis and M. E. Portnoi, {Tech. Phys. Lett.}
\textbf{31}, 671 (2005).

\bibitem{Kibis_07_NL} O. V. Kibis, M. Rosenau da Costa, and M. E.
Portnoi, {Nano Lett.} \textbf{7}, 3414 (2007).

\bibitem{Batrakov_06_SPIE}
K. G. Batrakov, P. P. Kuzhir, and S. A. Maksimenko, Proc. SPIE
\textbf{6328}, 63280Z (2006).

\bibitem{Kuzhir_07_SRIMN} P. Kuzhir, K. Batrakov, S. Maksimenko,
Synthesis and Reactivity in Inorganic, Metal-Organic and Nano-Metal
Chemistry \textbf{37}, 341 (2007)

\bibitem{Batrakov_08_PhE} K. G. Batrakov, P. P. Kuzhir, and S. A. Maksimenko, Physica E \textbf{40}, 1065 (2008).



\bibitem{Miano_06} G. Miano and F. Villone, IEEE Trans. Antennas Propagat.
\textbf{54}, 2713 (2006).

\bibitem {Hanson06}  J. Hao and G. W. Hanson, {Phys. Rev. B} \textbf{74}, 035119 (2006).

\bibitem {Hanson07}  J. Hao and G. W. Hanson, {Phys. Rev. B} \textbf{75}, 165416 (2007).

\bibitem {Shuba07}  M. V. Shuba, S. A. Maksimenko, and A. Lakhtakia, {Phys. Rev. B} \textbf{76}, 155407 (2007).

\bibitem{Huang_08} Y. Huang, W.-Y. Yin,  and Q. H. Liu, IEEE Trans.
Nano\-technol. \textbf{7}, 331 (2008).

\bibitem{Lan_06}Y. Lan, B. Zeng, H. Zhang, B. Chen, and Z. Yang, Int. J. Infrared Millim.
Waves  \textbf{27}, 871 (2006).

\bibitem {Bandow}  S. Bandow, M. Takizawa, K. Hirahara, M. Yudasaka, and S.
Iijima,  {Chem. Phys. Lett.} \textbf{337}, 48 (2001).

\bibitem {Iijima}  S. Iijima,  {Nature (London)} \textbf{354}, 56 (1991).

\bibitem {Ge}  M. Ge and K. Sattler,  {Science} \textbf{260}, 515 (1993).

\bibitem {wang05}  S. Wang and M. Grifoni,  {Phys. Rev. Lett.} \textbf{95}, 266802 (2005).

\bibitem {Saito93}  R. Saito, G. Dresselhaus, and M. S. Dresselhaus,  {Appl. Phys.} \textbf{73}, 494 (1993).

\bibitem {Lambin00}  P. Lambin, V. Meunier, and A. Rubio,  {Phys. Rev. B} \textbf{62}, 5129 (2000).

\bibitem {Ahn03}  K.-H. Ahn,Y.-H. Kim, J. Wiersig,
and K. J. Chang, {Phys. Rev. Lett.} \textbf{90}, 026601 (2003).

\bibitem {Yoon02}  Y.-G. Yoon, P. Delaney, and S. G. Louie, {Phys. Rev. B} \textbf{66}, 073407 (2002).

\bibitem {Lunde05}  A. M. Lunde, K. Flensberg, and A.-P. Jauho, {Phys. Rev. B} \textbf{71}, 125408 (2005).

\bibitem {Ho}  Y. H. Ho, G. W. Ho, S. C. Chen, J. H. Ho, and M. F. Lin, {Phys. Rev. B} \textbf{76}, 115422 (2007).

\bibitem {Kwon98}  Y.-K. Kwon and D. Tomanek, {Phys. Rev. B} \textbf{58}, R16001 (1998).

\bibitem {Abrikosov}  A. A. Abrikosov, D. V. Livanov, and A. A. Varlamov, {Phys. Rev. B} \textbf{71}, 165423 (2005).

\bibitem {Dyachkov} P. N. Dyachkov and D. V. Makaev, {Phys. Rev. B} \textbf{74}, 155442 (2006).

\bibitem {Tunney} M. A. Tunney and N. R. Cooper, {Phys. Rev. B} \textbf{74}, 075406 (2006).

\bibitem {Bourlon} B. Bourlon, C. Miko, L. Forro, D. C. Glattli, and A. Bachtold, Phys. Rev. Lett.  \textbf{93}, 176806 (2004).


\bibitem{balanis} C. A. Balanis, \textit{Antenna Theory: Analysis and Design\/} (Wiley, New York, NY, USA, 1997).

\bibitem{MKSK} S. A. Maksimenko, A. A. Khrushchinsky, G. Ya. Slepyan, and O. V. Kibis,
J. Nanophoton. {\bf 1}, 013505 (2007).


\bibitem{Ilyinsky} A. S. Ilyinsky, G. Ya. Slepyan, and A. Ya. Slepyan, \textit{Propagation,
Scattering and Dissipation of Electromagnetic Waves\/} (Peter
Peregrinus, London,  United Kingdom, 1993).

\bibitem{Novotny} L. Novotny and  B. Hecht, \textit{Principles of Nano-optics\/} (Cambridge University Press, Cambridge,
United Kingdom, 2006).

\bibitem{Colton} D. Colton and R. Kress, \textit{Integral Equation Methods in Scattering Theory\/}
(Wiley, New York, NY, USA, 1983).

\bibitem{Wolf} E. Wolf, Astrophys. Space Sci. {\bf 227}, 277 (1995).

\bibitem{Born} M. Born and E. Wolf, \textit{Principles of Optics\/}
(Pergamon, Oxford, 1999).

\bibitem{King} R. W. P. King and  T. T. Wu, \textit{IEEE
Trans. Antennas Propagat.} {\bf  14}, 524 (1966).

\bibitem {Lee} C.-K. Lee, J. Cho, J. Ihm, and K.-H. Ahn, {Phys. Rev. B} \textbf{69}, 205403 (2004).

\bibitem {Shuy} F. L. Shyu and M. F. Lin, Phys. Rev. B \textbf{62}, 8508
(2000).


\bibitem{Pitarke} J. M. Pitarke, V. M. Silkin, E. V. Chulkov, and P. M. Echenique, {Rep. Prog. Phys.} \textbf{70}, 1 (2007).

\bibitem {Novotny07} L. Novotny, {Phys. Rev. Lett.} \textbf{98}, 266802 (2007).

\bibitem{Elias} A. L. Elias, J. A. Rodriguez-Manzo, M. R. McCartney, D.
Golberg, A. Zamudio, S. E. Baltazar, F. Lopez-Urias, E.
Munoz-Sandoval, L. Gu, C. C. Tang, D. J. Smith, Y. Bando, H.
Terrones, and M. Terrones, {Nano Lett.} \textbf{5}, 467 (2005).

\bibitem {Hanson061} G. W. Hanson, {IEEE Trans. Antennas Propagat.} \textbf{54}, 76 (2006).

\bibitem{Johnson} P. B. Johnson and R. W. Christy, {Phys. Rev. B} \textbf{12}, 4370 (1972).

\bibitem {Wang01} B. Wang, S. Yin, G. Wang, A. Buldum,
and J. Zhao, {Phys. Rev. Lett.} \textbf{86}, 2046 (2001).

\bibitem{chin} K. C. Chin, A. Gohel, W. Z. Chen, H. I. Elim, W. Ji,
G. L. Chong, C. H. Sow, and A. T. S. Wee, {Chem. Phys. Lett.}
\textbf{409}, 85 (2005).

\bibitem{Lakht96} {\em Selected Papers on Linear
Optical Composite Materials\/} (A. Lakhtakia, ed.),  (SPIE Press,
Bellingham, WA, 1996).



\end{thebibliography}
\end{document}